\documentclass[useAMS,usenatbib,usegraphicx]{mn2e}
\usepackage{pdflscape}
\usepackage{multirow}
\usepackage{natbib}
\usepackage{aas_macros}
\usepackage{amsmath}

\title[Superyoung bursts in low-mass galaxies]{Detection of young ($\leq$20\,Myr) stellar 
populations in apparently quenched low-mass galaxies using red spectral line indices.}
\author[A. de Lorenzo-C\'aceres et al.]{A. de Lorenzo-C\'aceres$^{1,2}$\thanks{E-mail:
adrianadelorenzocaceres@gmail.com}, A. Vazdekis$^{1,2}$, J. Falc\'on-Barroso$^{1,2}$, M. A. Beasley$^{1,2}$
\\
$^{1}$Instituto de Astrof\'isica de Canarias, Calle V\'ia L\'actea s/n, E-38205 La Laguna, Tenerife, Spain\\
$^{2}$Departamento de Astrof\'isica, Universidad de La Laguna, E-38200 La Laguna, Tenerife, Spain\\
}
\begin{document}

\date{Accepted ***. Received ***; in original form ***}

\pagerange{\pageref{firstpage}--\pageref{lastpage}} \pubyear{2002}

\maketitle

\label{firstpage}

\begin{abstract}
We report on the detection of a small contribution (around and below 1\% in mass) from young stellar components with ages
$\leq$20\,Myr in low-mass galaxies purposely selected from the MaNGA survey to be already-quenched systems.
Among the sample of 28 galaxies, 
eight of them show signatures of having suffered a very recent burst of star formation.
The detection has been done through the analysis of line-strength indices in the red spectral range
$[$5700,8800$]$\,\AA.
The increasing contribution of red supergiants to this red regime is responsible for a deviation
of the index measurements with respect to their position within the model grids in the
standard spectral range $[$3600,5700$]$\,\AA. 
We demonstrate that a combination of red indices, as well as a qualitative assessment of the 
mean luminosity-weighted underlying stellar population, is required in order to distinguish
between a true superyoung population and other possible causes of this deviation, such as abundance ratio variations.
Our result implies that many presumably quenched low-mass galaxies actually contain gas that is triggering
some level of star formation. They have, therefore, either accreted external gas, internally recycled enough gas from
stellar evolution to trigger new star formation, or they kept a gas reservoir after
the harassment or stripping process that quenched them in first place. 
Internal processes 
are favoured since we find no particular trends between our non-quenched galaxies and their environment, although
more work is needed to fully discard an external influence.
\end{abstract}


\section{Introduction}\label{sec:intro}
The absorption lines in the spectrum of a galaxy contain information about its stellar populations.
Measurements of such lines, the so-called line-strength indices, 
have been widely used in the literature to analyse the mean age, metallicity, and relative abundance of 
chemical elements of stars in galaxies 
\citep[e.g. ][]{Trageretal98, Spolaoretal2010,deLorenzoCaceresetal2013} 
and, particularly, in low-mass objects 
\citep[e.g.][]{Gorgasetal1997,Michielsenetal2003,Senetal2018,Sybilskaetal2018}. 
This is done by comparison with
predictions from synthesis models of single-age, single-metallicity stellar populations (SSP),
such as those from \citet{BruzualandCharlot2003} and
\citet[][]{Vazdekisetal2010}.
The analysis of line-strength indices
constituted a milestone for the study of stellar content in galaxies, which 
nowadays has been complemented by full spectrum fitting techniques that also make use
of SSP models as fitting templates. As examples, see \citet{delaRosaetal2016}, \citet{Sanchezetal2018},
\citet{deLorenzoCaceresetal2012} and \citet{deLorenzoCaceresetal2019b}
for the application of different full spectrum fitting codes to observed galaxies, namely 
\texttt{STARLIGHT} \citep{CidFernandesetal2005}, \texttt{Pipe3D} \citep{Sanchezetal2016}, \texttt{ULySS}
\citep{Kolevaetal2009} and
\texttt{STECMAP} \citep{Ocvirketal2006b}, respectively.\\

Due to the large amount of data available, the behaviour of stellar populations studied through 
line-strength indices in the optical spectral range is 
well known: massive ellipticals show old age ($\sim$12\,Gyr for redshift $z=$0), 
solar or larger metallicity, 
and enhanced $[$Mg/Fe$]$ and $[$Na/Fe$]$ values up to 0.5\,dex 
\citep[e.g.][]{Kuntschner2000, LaBarberaetal2013, LaBarberaetal2017}. Indeed, 
a correlation between stellar population parameters and galaxy mass is found for all massive galaxies,
with a moderate dependance on the environment 
\citep[e.g. ][]{Carreteroetal2004, Carreteroetal2007, Onoderaetal2015}. Note however that
very small fractions (below 1\% in mass) of young stars have recently been detected in 
red massive galaxies at 0.35$<$$z$$<$0.6, thus suggesting that star formation is not completely
halted in such old systems 
\citep{SalvadorRusinoletal2020}.

The age-mass correlation holds in the low-mass regime \citep{Kolevaetal2011}: 
in general, dwarfs are found to be younger and
less metal-rich than their massive counterparts \citep[see][among others]{HeldandMould1994,Rakosetal2001}.
These results are indicative of the downsizing scenario, meaning that low-mass galaxies
have more extended star formation histories than massive ellipticals, as further supported by
their $[\alpha$/Fe$]$ values around solar \citep{Gorgasetal1997, Michielsenetal2007}.
Numerical simulations also obtain similar results, at least for the most massive dwarfs
with M$\sim$10$^9$M$_\odot$ \citep{GarrisonKimmeletal2019}.

The role of the environment in regulating star formation in low-mass galaxies is controversial:
whereas it seems reasonable that star formation in cluster dwarfs is halted through very efficient
ram-pressure stripping when entering high-density environments, 
robust observational evidence of such quenching has not been found yet.
For example, \citet{Michielsenetal2008} analysed a sample of 18 dwarfs in the Virgo cluster,
finding that galaxies close to the cluster core are older than those at higher Virgocentric distances.
On the contrary, \citet{Kolevaetal2009b}
did not find any significant connection between the environment and star formation histories
of dwarfs in the Fornax cluster and in groups.

More recently, \citet{Senetal2018} analysed a sample of
37 dwarf ellipticals in the Virgo cluster, for which an age range between 1\,Gyr and 10\,Gyr and 
a metallicity range $[$Fe/H$]\epsilon[$-1.,0.$]$\,dex were measured. 
These authors also explored the abundance ratios, finding that $[$Mg/Fe$]$ 
acquires close to solar values as expected, whereas $[$Na/Fe$]$
is clearly underabundant with respect to the solar neighbourhood.
In the context of the hELENa project, \citet{Sybilskaetal2017, Sybilskaetal2018}
studied the stellar content of 
20 dwarf ellipticals in different environments (17 targets in various 
regions of the Virgo cluster, the remaining 3 galaxies in the field),
finding consistent results with those presented in \citet{Senetal2018}. While metallicity 
and $\alpha$-enhancement do not vary with galaxy environment, dwarf ellipticals appear to be 
younger when they live in low-density regions, in agreement with \citet{Michielsenetal2008}.  
\citet{Sybilskaetal2017, Sybilskaetal2018} concluded that low-mass objects
come from late-type analogous whose gas was stripped during their trip towards 
dense environments.  

In summary, the evolution of dwarfs and its dependence on the environment still pose a complicated picture,
and the downsizing connection between massive ellipticals and dwarfs remains under debate.
Caution is required when comparing galaxies through their mean stellar populations. 
\citet{Rysetal2015} showed that some dwarf ellipticals host young stars whose contribution alters the mean luminosity-weighted
age of an otherwise old system. Whether allegedly quenched dwarf galaxies may be going through some star formation 
process, in an analogous way as 
massive early-type galaxies do \citep[see recent results by][using the predictions from \citealt{Vazdekisetal2016}
for the UV spectral range]{SalvadorRusinoletal2020},
is therefore a fundamental question that we address in this paper.\\

Observationally, the measurements of line-strengh indices for SSPs with ages older than
1\,Gyr are well behaved. 
For example, the H$\beta_{\rm O}$ index
defined by \citet{CervantesandVazdekis2009} acquires decreasing values from $\sim$6\,\AA\ to $\sim$2\,\AA\ in that age range, 
H$\delta_{\rm F}$ \citep{WortheyandOttaviani97} spans from $\sim$4\,\AA\ to $\sim$0\,\AA, and
$[$MgFe$]'$ \citep{Thomasetal2003} increases from $\sim$1\,\AA\ to $\sim$3\,\AA.
Such well-behaved trend is however lost when younger stellar populations are considered:
the Balmer age-sensitive indices show 
a turn-off point peaking at approximately 300\,Myr 
\citep[][]{Frogeletal1990}, 
with their values decaying again
for even younger SSPs. Regarding the metallicity-sensitive indices, 
they continue their monotonically decrease as in the old regime
until 20\,Myr-old SSPs are reached. The contribution from supergiant stars is then
maximum, when the red supergiants (RSGs) are particularly 
influential. RSGs indeed make some metallicity-sensitive species, such as Ca or TiO,
show a local (even absolute in some cases) peak at $\sim$10\,Myr \citep{Asadetal2017}. 
The redder the spectral range, the larger the effect of RSGs
on the absorption lines and the more prominent the peak.\\

In this work, we take advantage of the noticeable non-linearity of the behaviour 
of absorption line-strength indices in the red wavelength range. We 
report the detection of a small ($\leq$1\% in mass) contribution from stars younger than
20\,Myr in eight low-mass galaxies from the MaNGA survey previously identified as quenched systems.
The selection of the total sample of 28 low-mass galaxies is described in Sec.\,\ref{sec:sample}.
The measurement of the absorption line-strength indices and their behaviour, including 
some analyses aiming at constraining the properties of the young component as much as possible,
are explained in Sec.\,\ref{sec:analysis} and \ref{sec:results}.
The capabilities and caveats of the use of the red wavelength regime for detecting 
such young populations, which we refer to as ``superyoung'', are discussed in Sec.\,\ref{sec:discussion},
where we also discuss the implications of our findings for the evolution of low-mass galaxies.

\section{Low-mass galaxies from the MaNGA survey}\label{sec:sample}
\begin{table*}
 \centering
  \caption{Properties of the final sample of 28 low-mass galaxies: MaNGA identification
number; stellar mass and redshift
as included in the NASA Sloan Atlas catalog (www.nsatlas.org); number of spaxels
used in the stacking (see text in Sec.\,\ref{sec:analysis}); SNR (calculated in the red
continuum of H$\beta$) of the final stacked spectrum for each galaxy; and median of the emission flux
in H$\alpha$ in the same spaxels used for the stacking, as calculated through a Gaussian fitting 
with the MaNGA Data Analysis Pipeline \citep{Westfalletal2019, Belfioreetal2019}. The error value corresponds to the standard
deviation of the H$\alpha$ flux in the same spaxels.
The stellar masses shown here have been calculated by using the elliptical Petrosian 
fluxes and a flat cosmology with $\Omega_{m}=$0.3, $\Omega_{\Lambda}=$0.7, and $H_0=$100\,km\,s$^{-1}$\,Mpc$^{-1}$.
Galaxies with detected $\leq20$\,Myr stellar contributions are highlighted in bold face.}
  \label{tab:sample}
  \begin{tabular}{cccccc}
  \hline
MaNGA ID & Stellar mass & Redshift & \# spaxels & SNR & Flux(H$\alpha$)\\
 &  (M$_{\odot}$) & & & & erg/s/cm$^2$/spaxel\\
\hline
\hline
{\bf 1-217044}  & {\bf 2.53$\times$10$^{9}$}  &  {\bf 0.027}  &  {\bf 45}  &  {\bf 36.2} & {\bf 0.045 $\pm$ 0.045}\\
{\bf 1-92638 }  & {\bf 2.66$\times$10$^{9}$}  &  {\bf 0.038}  &  {\bf 69}  &  {\bf 37.1} & {\bf 0.082 $\pm$ 0.088}\\
{\bf 1-38157 }  & {\bf 2.72$\times$10$^{9}$}  &  {\bf 0.038}  &  {\bf 38}  &  {\bf 30.4} & {\bf 0.074 $\pm$ 0.140}\\
{\bf 1-256457}  & {\bf 2.90$\times$10$^{9}$}  &  {\bf 0.037}  &  {\bf 81}  &  {\bf 36.9} & {\bf 0.175 $\pm$ 0.086}\\
{\bf 1-211098}  & {\bf 2.93$\times$10$^{9}$}  &  {\bf 0.028}  &  {\bf 32}  &  {\bf 27.7} & {\bf 0.070 $\pm$ 0.198}\\
{\bf 1-256125}  & {\bf 3.09$\times$10$^{9}$}  &  {\bf 0.038}  &  {\bf 94}  &  {\bf 38.0} & {\bf 0.154 $\pm$ 0.198}\\
{\bf 12-49536}  & {\bf 3.23$\times$10$^{9}$}  &  {\bf 0.019}  &  {\bf 34}  &  {\bf 56.9} & {\bf 0.045 $\pm$ 0.065}\\
{\bf 1-255220}  & {\bf 3.54$\times$10$^{9}$}  &  {\bf 0.022}  & {\bf 300}  &  {\bf 49.4} & {\bf 0.143 $\pm$ 0.184}\vspace{0.1cm}\\
1-258746  & 1.17$\times$10$^{9}$  &  0.024  &  38  &  33.5  &  0.079 $\pm$ 0.080\\
1-277462  & 1.29$\times$10$^{9}$  &  0.022  &  40  &  55.4  &  0.116 $\pm$ 0.140\\
1-519705  & 1.75$\times$10$^{9}$  &  0.028  &  96  &  25.0  &  0.150 $\pm$ 0.575\\
1-252147  & 2.51$\times$10$^{9}$  &  0.018  &  249 &  39.5  &  0.130 $\pm$ 0.330\\
1-94958   & 2.52$\times$10$^{9}$  &  0.034  &  96  &  50.5  &  0.098 $\pm$ 0.292\\
1-38319   & 2.70$\times$10$^{9}$  &  0.038  &  93  &  29.4  &  0.058 $\pm$ 0.112\\
1-113520  & 2.76$\times$10$^{9}$  &  0.017  &  294 &  53.4  &  0.244 $\pm$ 0.333\\
1-133948  & 2.77$\times$10$^{9}$  &  0.019  &  42  &  52.3  &  0.085 $\pm$ 0.166\\
1-43679   & 2.94$\times$10$^{9}$  &  0.029  &  179 &  32.5  &  0.038 $\pm$ 0.050\\
1-277159  & 3.31$\times$10$^{9}$  &  0.025  &  99  &  43.2  &  0.040 $\pm$ 0.061\\
1-209113  & 3.48$\times$10$^{9}$  &  0.038  &  82  &  32.2  &  0.097 $\pm$ 1.273\\
1-322087  & 3.62$\times$10$^{9}$  &  0.039  &  109 &  28.7  &  0.054 $\pm$ 0.232\\
1-211044  & 3.72$\times$10$^{9}$  &  0.028  &  103 &  37.7  &  0.113 $\pm$ 0.474\\
1-567184  & 4.02$\times$10$^{9}$  &  0.025  &  180 &  40.9  &  0.082 $\pm$ 0.095\\
1-115062  & 4.17$\times$10$^{9}$  &  0.026  &  225 &  31.2  &  0.162 $\pm$ 0.266\\
12-110746 & 4.30$\times$10$^{9}$  &  0.029  &  160 &  30.2  &  0.114 $\pm$ 1.111\\
1-634477  & 4.35$\times$10$^{9}$  &  0.023  &  247 &  47.0  &  0.062 $\pm$ 0.134\\
1-322680  & 4.56$\times$10$^{9}$  &  0.037  &  143 &  30.7  &  0.075 $\pm$ 3.369\\
1-211019  & 4.88$\times$10$^{9}$  &  0.030  &  212 &  32.2  &  0.125 $\pm$ 0.852\\
1-629695  & 5.17$\times$10$^{9}$  &  0.028  &  280 &  52.8  &  0.181 $\pm$ 0.185\\
\hline       
\end{tabular}
\end{table*}

\begin{figure*}
 \vspace{2pt}
 \includegraphics[bb= 54 30 530 230, angle=0., width=1.\textwidth]{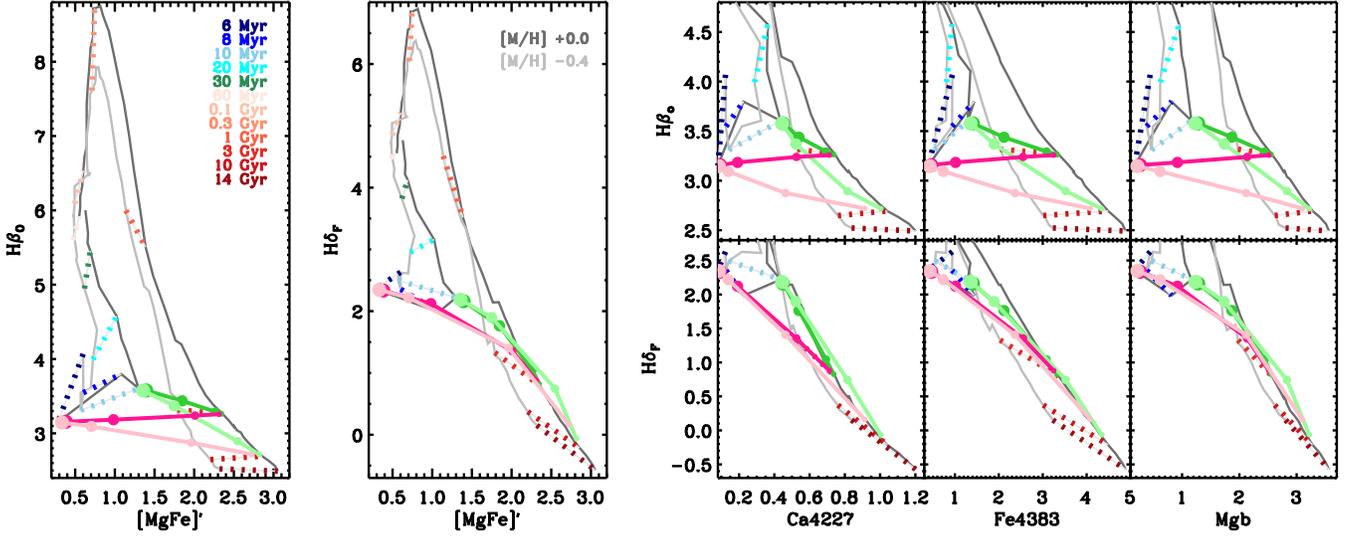}
 \caption{Standard indices in the optical range. The vertical axes show 
age-sensitive indices: H$\beta_{\rm o}$ and H$\delta_{\rm F}$. 
The horizontal axes show metallicity-sensitive indices; from left to right: 
$[$MgFe$]'$, Ca4227, Fe4383, and Mg$_b$. For the sake of clarity, the 
full dynamical range for the age-sensitivy indices is shown against $[$MgFe$]'$, whereas
we perform a zoom around the bottom regions of interest for the remaining diagrams.
The grids correspond to predictions from the E-MILES models at LIS-14\,\AA, 
with ages from 6\,Myr to 14\,Gyr (colour-coded as indicated by the legend) and metallicity 
$[$M/H$]$=-0.4 and $[$M/H$]$=0.0 (light and dark grey lines, respectively). 
The circles track the behaviour for a combination of: i) a single stellar population model with solar metallicity
and age of 10\,Gyr (light colours) and 3\,Gyr (dark colours); and ii) a 0.01\%-0.1\%-1\%-10\% varying
mass fraction (increasing as the increasing size of the circles) 
of a young component with age 6\,Myr (pink) and 10\,Myr (green). Only superyoung mass fractions
significantly larger than 0.1\% (1\% for 10\,Myr-old superyoung populations) 
cause a noticeable effect on these index-index measurements.} 
 \label{fig:standardSSP}
\end{figure*}
\begin{figure*}
 \vspace{2pt}
 \includegraphics[bb= 54 30 600 230,angle=0., width=1.\textwidth]{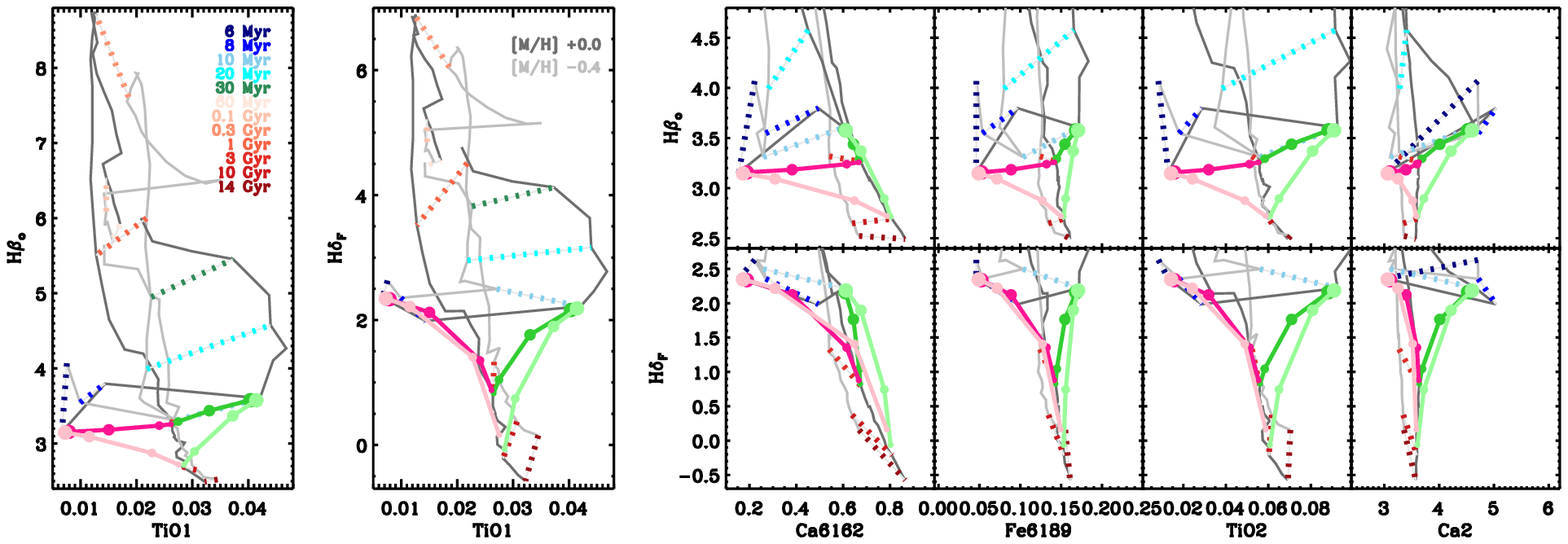}
 \caption{Same as Fig.\,\ref{fig:standardSSP} but for redder metallicity-sensitive indices:
TiO$_1$, Ca6162, Fe6189, TiO$_2$, and Ca2 (from left to right). Note that the 10\,Myr SSP prediction
lays below the dark green tracking line. While the SSP model grids do not allow to constrain 
the age and metallicity of an old population, superyoung mass fractions even below 1\% shift the measurements
out of the SSP predictions and in opposite directions depending on their age.}
 \label{fig:redSSP}
\end{figure*}

The low-mass galaxies under study here belong to the sample
of 39 objects that \citet{Pennyetal2016} withdrew from the MaNGA multi-object integral-field spectroscopic
survey \citep{Bundyetal2015,Wakeetal2017}, observed with the Baryon Oscillation 
Spectroscopic Survey spectrograph \citep[BOSS; ][]{Smeeetal2013}. MaNGA has
been conceived as part of the Sloan Digital Sky Survey (SDSS)-IV main project
with the goal of observing $\sim$10\,000 galaxies with masses higher than 
5$\times$10$^8$\,M$_\odot$ \citep{Blantonetal2017}. BOSS provides spectra covering
$[$3600, 10\,300$]$\,\AA\, with a  resolution of $\sigma\sim$77\,km\,s$^{-1}$
(mean value, as the actual resolution is variable with wavelength),
a spectral sampling of $\Delta(log\,\lambda)=10^{-4}$,
and a spatial sampling of 0.5\,arcsec per spaxel after data reduction.

\citet{Pennyetal2016} selected galaxies with magnitudes M$_{\rm r}>$-19 and 
masses ranging from 1$\times$10$^9$\,M$_\odot$
to 5$\times$10$^9$\,M$_\odot$. They purposely picked up quenched galaxies,
as assessed through two tracers of current star formation, namely
weak H$\alpha$ equivalent widths in emission (EW$_{\rm H\alpha}<$2\,\AA) and red colours ($u-r$)$>$1.9. 
The subtle star formation bursts we aim at detecting have a larger effect on low-mass
galaxies than on massive galaxies, where they are absolutely negligible due to the 
little relative contribution of light from these young bursts with respect to the
total galaxy light \citep[see notwithstanding][]{SalvadorRusinoletal2020}.
Quenched galaxies are also required for carrying out this project, since 
they populate the oldest regions of the absorption line-strength index-index
diagrams \citep{Gehaetal2012}. 
This fact maximises the contrast between the superyoung contribution and
the underlying stellar population, and makes deviations towards younger regions of the 
diagrams noticeable. Moreover, the contribution of recent starbursts 
to an extended star formation history gets diluted, whereas their effect over an
old single-burst-like spectrum should be easier to detect even in the case of 
very small mass fractions \citep[see][]{Vazdekisetal2016}.\\

The data reduction of the MaNGA spectra 
is performed with a dedicated pipeline presented in
\citet{Lawetal2016}. Some data analysis is also provided by the MaNGA collaboration
following the procedure presented in \citet{Westfalletal2019, Belfioreetal2019}.
This includes measurement of the stellar kinematics, emission lines,
and various line-strength spectral indices. In particular, the stellar kinematics is 
fitted by applying the \texttt{pPXF} code \citep{CappellariandEmsellem2004} 
in combination
with the MILES empirical stellar library from 
\citet[][see also \citealt{FalconBarrosoetal2011}]{SanchezBlazquezetal2006}.
In this work we have made use of the
pure stellar spectra (i.e. after removal of the weak emission lines) and stellar kinematics
provided by
the MPL-5 data release of the analysed MaNGA spectra. Note that
due to the quenched nature of our sample galaxies, little if not negligible
ionised gas content is present. 

Our analysis, explained in Sec.\,\ref{sec:analysis}, includes an initial stacking process to increase the 
signal-to-noise (SNR) of each galaxy spectrum. 
From the original sample of 39 galaxies, only individuals with more than 30 spaxels
with SNR$\geq$10 and flagged as with no problems during the reduction and stellar kinematic analysis are considered
(see more details about these processes in Sec.\,\ref{sec:analysis}). 
After this refinement, our final sample comprises 28 low-mass galaxies, whose main properties (MaNGA identification
number, redshift, and mass) are shown in Table\,\ref{tab:sample}, together with the SNR in the
red continuum of H$\beta$ of the final stacked spectrum
and the number of spaxels involved in the stacking.
Mass and redshift are retrieved from the NASA Sloan Atlas catalog\footnote{www.nsatlas.org} and
the stellar masses in particular have been calculated by using the elliptical Petrosian 
fluxes and a flat cosmology with $\Omega_{m}=$0.3, $\Omega_{\Lambda}=$0.7, and 
$H_0=$100\,km\,s$^{-1}$\,Mpc$^{-1}$.\\

\begin{figure*}
 \vspace{2pt}
 \includegraphics[bb= 54 30 530 230, angle=0., width=1.\textwidth]{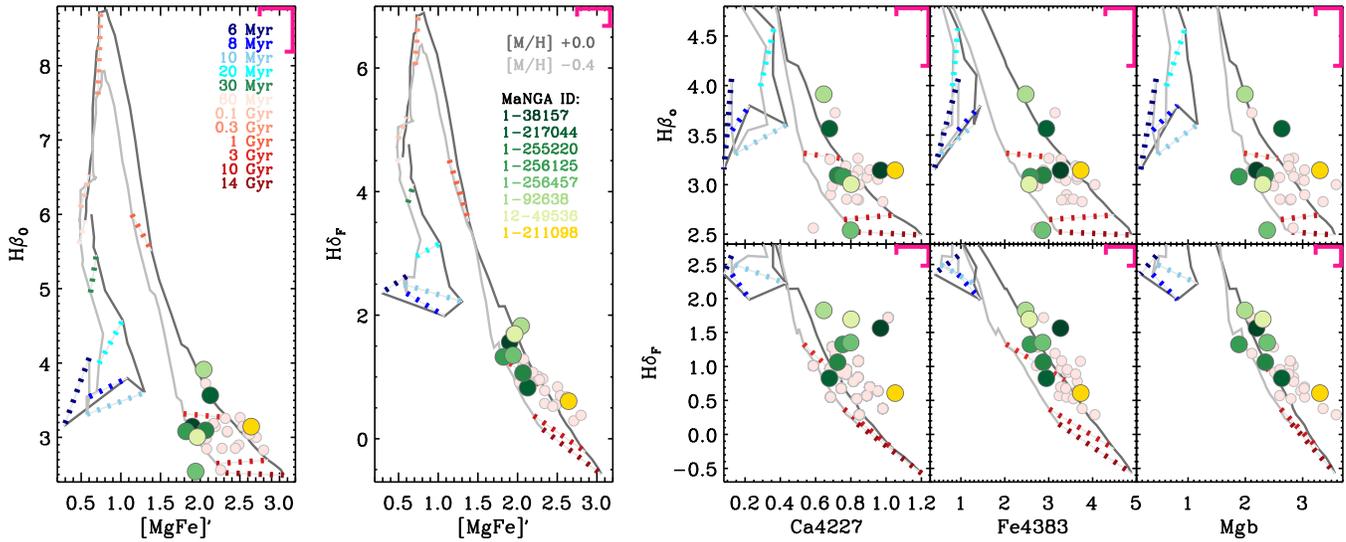}
 \caption{Line-strength indices and E-MILES model predictions as in Fig.\,\ref{fig:standardSSP}.
Measurements for the low-mass galaxies are plotted with pale pink circles; green-coloured circles correspond 
to the eight out of 28 galaxies for which a recent burst of star formation is detected, 
the yellow circle being the galaxy used for explanations in Sec.\,\ref{sec:red} 
(see text for details). The MaNGA identification numbers of the selected galaxies are indicated in the
H$\delta_{\rm F}$ vs. $[$MgFe$]'$ panel. Mean error bars are shown in the top right corner of each 
panel (in magenta).}
 \label{fig:standard}
\end{figure*}

\begin{figure*}
 \vspace{2pt}
 \includegraphics[bb= 54 30 600 230,angle=0., width=1.\textwidth]{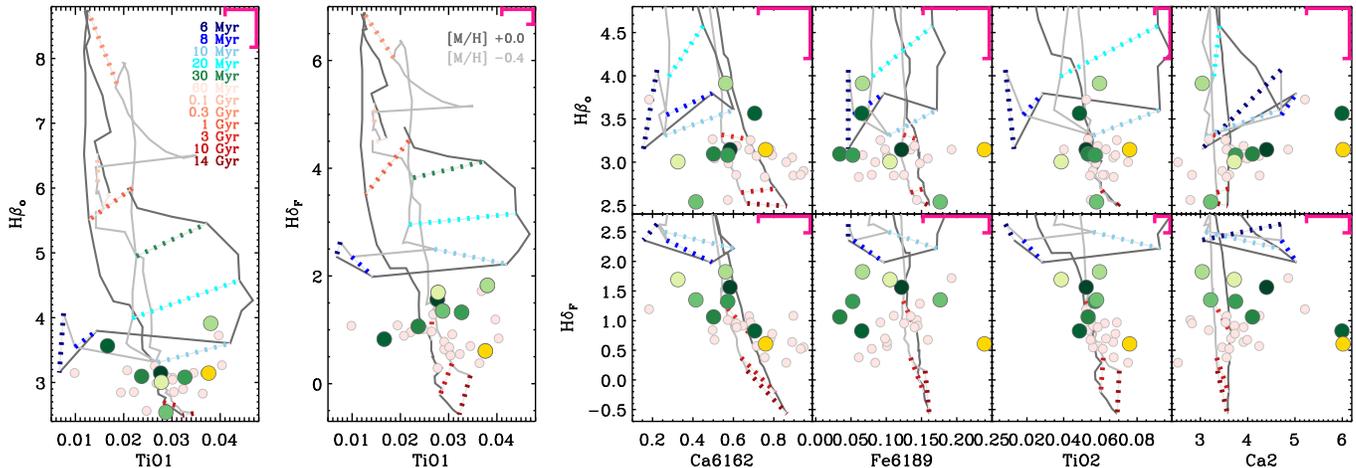}
 \caption{Line-strength indices and E-MILES model predictions as in Fig.\,\ref{fig:redSSP}.
Measurements for the low-mass galaxies are plotted with pale pink circles; green-coloured circles correspond 
to the eight out of 28 galaxies for which a recent burst of star formation is detected since they are 
shifted in all panels in a way consistent with hosting a slight fraction of a superyoung stellar population. 
The yellow circle corresponds to the galaxy used for explanations in Sec.\,\ref{sec:red} 
(see text for details). Mean error bars are shown in the top right corner of each 
panel (in magenta).}
 \label{fig:red}
\end{figure*}

\section{Analysis and measurement of line-strength indices}\label{sec:analysis}
In order to maximise the signal-to-noise and the likelihood
of detecting very recent bursts which might have happened anywhere 
in the galaxy, our analysis starts by 
stacking all galaxy spectra within the MaNGA integra-field unit.
The low-mass galaxies have been observed with the 19-fibres bundle from which we
select the spectra with SNR$\geq$10 and flagged as with no problems
during the reduction process. A good quality measurement of the stellar velocity
and velocity dispersion is also required and assessed through the corresponding
flags in the MPL-5 analysis. The measurement of the stellar velocity dispersion
is particularly restrictive. By selecting only those spectra of the highest quality
we sometimes end up with few elements for the stacking. 
Due to this, we only keep galaxies with more than 30 spectra suitable for the 
stacking in the final sample selection, as explained in Sec.\,\ref{sec:sample}.

Before stacking, all spectra are (i) shifted to a common velocity selected as the
systemic velocity of the galaxy. We note that each spectrum is shifted 
a whole number of spectral elements
($\Delta(log\,\lambda)=10^{-4}$) to prevent biases due to interpolation. Such strategy
is further supported by the fact that the uncertainty corresponds to $\leq$1\% of the galaxy redshift; 
(ii) normalised to the same flux intensity in the red continuum of
H$\beta$, i.e., $[$4870, 5000$]$\,\AA. This means that the outer galaxy regions
are considered at the same level that the brightest, central parts; and (iii) 
degraded to a common and constant-with-$\lambda$ velocity dispersion
corresponding to a full-width-at-half-maximum FWHM$=$14\,\AA. 
Such step is done in order to work in the LIS-14.0\,\AA\ system defined by 
\citet{Vazdekisetal2010}, for which predictions of the line-strength indices
measured on the single stellar population models based on the
MILES spectral library \citep{SanchezBlazquezetal2006, FalconBarrosoetal2011}
are provided.\\

Line-strength indices are measured over the stacked spectra for the 28 
low-mass galaxies under study. As age-sensitive indicators, 
we use the H$\beta_{\rm o}$ and H$\delta_{\rm F}$ indices defined by
\citet{CervantesandVazdekis2009} and \citet{WortheyandOttaviani97}, respectively.
Among the metallicity-sensitive indicators, we refer to Ca4227,
Fe4383 \citep[both redefined in ][]{Trageretal98}, Mg$b$ \citep{Wortheyetal94}, 
and $[$MgFe$]'$ \citep{Thomasetal2003} as \emph{standard} indices.
We call \emph{red} metallicity-sensitive indices those that belong to the
spectral range between 5700\,\AA\ and 8800\,\AA. In particular, we use
TiO$_1$, TiO$_2$ \citep{Gregg94,Trageretal98}, 
Ca6162, Fe6189, and the Ca2 index which maps the central line of the Calcium Triplet
\citep{Cenarroetal2001}.\\

Galaxy indices are compared with the same measurements over the
MILES-based synthesis stellar population models of 
\citet[][]{Vazdekisetal2016}.
They represent a wavelength-extended version (E-MILES) of the base models of 
\citet{Vazdekisetal2010}, which are mostly solar scaled at solar metallicity. The E-MILES models cover the 
whole spectral range from 1680\,\AA\ to 50\,000\,\AA\ with a 
wavelength-dependent resolution which acquires the constant value
of FWHM$=$2.51\,\AA\ in the spectral range we are interested in, i.e., 
from 3600\,\AA\ to 8800\,\AA. 

The E-MILES single stellar population models (SSPs) cover a range of metallicity from 
-1.79 to +0.26 and ages older than 63\,Myr. We have complemented them
with the analogous superyoung models\footnote{Models available at miles.iac.es} 
spanning down to 6\,Myr \citep[see details in][]{Asadetal2017}.
The SSPs used in this work
have been computed by means of the isochrones from \citet[][E-MILES]{Girardietal2000}
and \citet[][superyoung MILES]{Bertellietal94}, 
and assuming the
universal initial mass function of \citet{Kroupa2001}.
Figures\,\ref{fig:standardSSP} and \ref{fig:redSSP} show the model grids including predictions for
index values in the standard and red regimes, respectively. Note the
discontinuity observed for the SSP predictions between the E-MILES
and MILES superyoung models: there is a break in the metallicity lines of the grids at 60\,Myr that is
larger for solar metallicity than for $[$M/H$]$$=$-0.4, and for the red indices than for the standard indices. Such discontinuity 
is due to a mild mismatch between the 
isochrones that mainly affects those indices with large variations for ages
$<$30\,Myr. Differences are nevertheless low and they do not affect the results obtained in this work.

\subsection{Observing the effect of a superyoung contribution over an old single stellar population.}
Figures\,\ref{fig:standardSSP} and \ref{fig:redSSP} show the behaviour across the index-index 
diagrams of a combination of two single stellar populations with solar metallicity: an old SSP of 3\,Gyr or 10\,Gyr,
and a superyoung SSP of 6\,Myr or 10\,Myr, 
thus resembling a two-bursts star formation history.
The mass fraction of the superyoung component varies as 0.01\%, 0.1\%, 1\%, and 10\%.\\

For the standard indices (Fig.\,\ref{fig:standardSSP}), 
superyoung mass fractions below 1\% only cause a slight shift of the
index-index measurements, which remain within the area covered by the model grids. The equivalent
mean luminosity-weighted age and metallicity of this two-bursts star formation history is 
therefore sligthly altered (as younger and/or more metal poor, as expected according to the 
age-metallicity degeneracy reflected in the overall shape of the model grid), but the plots do not reveal the 
presence of a superyoung contribution. Indeed, only a mass fraction significantly larger than 0.1\% 
of a 6\,Myr-old component produces a shift of the measurements out of the model grids and towards the 
superyoung regime that could be noticeable by inspecting these standard plots.\\

Red metallicity-sensitive indices provide more information about superyoung contributions 
than the standard ones, as shown in Fig.\,\ref{fig:redSSP}. Given the shape of
the model grids, with narrow and almost vertical areas covered by SSPs older than 300\,Myr,
these indices are not suitable for determining the mean luminosity-weighted
age and metallicity of old populations. The model grids however widen for very young populations,
particularly younger than 20\,Myr, for which the area covered by the model predictions also shifts horizontally and
therefore does not overlap with the vertical region of the $>300$\,Myr populations.
Thus, this regime allows not only to detect very young contributions but also to constrain
the age of the superyoung component by combining different indices where the shifts are inverted.
Note how the tracking lines for our two-bursts star formation histories with 
6\,Myr and 10\,Myr-old superyoung contributions go in opposite directions. 

As indicated in Sec.\,\ref{sec:sample}, this kind of analysis requires the mean luminosity-weighted
age of the underlying population, whatever its star formation history is, to be old
(particularly older than 3\,Gyr), so the measurements lie in the bottom part of the index-index
diagrams where the trends can be observed cleanly. Fig.\,\ref{fig:redSSP} shows how superyoung contributions
as small as 0.1\% in mass can cause noticeable shifts in these diagrams. On the other end, 
when the superyoung population represents 10\% of the total galaxy mass, it is overwhelmingly dominant 
in light and therefore drives the measurements in both the standard and red regimes (see how the 
indices measurements overlap over the 6\,Myr and 10\,Myr SSP predictions
regardless of the age of the old underlying population in both Fig.\,\ref{fig:standardSSP} and \ref{fig:redSSP}).

\section{Results: a trip from the standard to a redder spectral regime}\label{sec:results}

\subsection{Standard indices agree with the sample galaxies being quenched and metal poor}
Figure\,\ref{fig:standard} shows the standard indices 
for the 28 low-mass galaxies of our sample.
The panels corresponding to the $[$MgFe$]'$ total metallicity indicator show how
all measurements are well constrained within the grid, with ages ranging from 
$\sim$2\,Gyr and 10\,Gyr and metallicity between solar and $[$M/H$]=$-0.4.
H$\beta_{\rm o}$ better breaks the age-metallicity degeneracy
and provides a more orthogonal model grid than 
H$\delta_{\rm F}$. However, emission in H$\beta$ is expected to be more prominent than
in H$\delta$ so, for the sake of completeness and even though we expect little 
gas in these allegedly quenched galaxies (see Sec.\,\ref{sec:sample}), we show both age-sensitivity
indices throughout the paper. Regarding metallicity, we warn that the tiny fractions of young populations
found in this work may be partially responsible for the relative differences in metallicity found
for some galaxies between the H$\beta_{\rm o}$ and H$\delta_{\rm F}$ diagrams. 

The low-mass galaxies show close to solar $[$Mg/Fe$]$ abundance ratio, 
in agreement with the findings of \citet{Senetal2018} for dwarf elliptical galaxies.
Such behaviour is shown by comparison of Z$_{\rm Mg\emph{b}}$ and Z$_{\rm Fe4383}$ 
\citep[see][among others, for a justification of the use of this proxy]{Vazdekisetal2001,Vazdekisetal2010}, 
taking into account that the E-MILES models follow the abundance pattern of the Milky Way and therefore
the behaviour of stars in the solar vicinity needs to be considered:
$[$M/H$]=$0.0 and Z$_{\rm Mg\emph{b}}=$Z$_{\rm Fe4383}$ correspond to $[$Mg/Fe$]=$0.0, whereas $[$M/H$]=$-0.4 and 
Z$_{\rm Mg\emph{b}}=$Z$_{\rm Fe4383}$ correspond to $[$Mg/Fe$]=$0.2-0.3.

Analogously, Fig.\,\ref{fig:standard} shows an enhanced $[$Ca/Fe$]$ ratio for the sample
galaxies in comparison with the solar value. This finding was also previously 
observed and discussed in \citet{Senetal2018} and it is opposed to what occurs in massive early-type
galaxies \citep{Vazdekisetal97}.

\subsection{Red indices in the R-, I-bands reveal the contribution of 
a $<$20\,Myr population}\label{sec:red}

Figure\,\ref{fig:red} shows what we refer to as red indices, from TiO$_1$ to the Calcium Triplet CaT
(see Sec\,\ref{sec:analysis}).
The data points for the low-mass galaxies, 
well constrained within the model grids of the standard indices,
spread out and populate \emph{forbidden} regions of the SSP-based diagrams.
Such shifts may be explained by abundance ratios and to a less extent by
the contribution of a small fraction of very young stars of ages $\leq$20\,Myr.
Note that potential contamination by telluric lines, particularly important for the
Fe6189, TiO$_1$, and TiO$_2$ indices (depending on the galaxy redshift),
has been purposely assessed and discarded.

After a careful visual exploration of all individual cases taking into account
the overabundance estimates provided by the standard indices, we have identified 
eight out of 28 low-mass galaxies for which shifts in all panels are compatible
with the presence of a superyoung stellar population. 
Those galaxies are highlighted with colours in Fig.\,\ref{fig:red},
as well as in Fig.\,\ref{fig:standard}. It is noticeable
that the data points have moved towards younger parts of the diagram with respect
to the mean luminosity-weighted measurements from the standard indices,
following a tilted path (both vertical and horizontal movement) 
towards the regions populated by $\leq$20\,Myr stars. 
The indices for which measurements of very young populations move unequivocally towards the left or right
regions of the diagram are the ones for which the selected eight galaxies show
a largest departure from the old stellar population regime of the grid. Such effect
can be seen in the Ca6162 (left shift) and Ca2 (right shift) panels; the fact that
the datapoints move in opposite directions for the same
element (Calcium) indicates abundance ratio is not the reason behind the shifts.
On the other hand, the indices for which the shifts for young populations younger and older than
$\sim$10\,Myr go in opposite directions (left and right inside the same index-index diagram), 
such as TiO$_1$ and TiO$_2$, show a 
large scatter of the cloud composed of the eight coloured circles.

Some of the remaining twenty galaxies shift with respect to their standard
properties as well, but not in a consistent manner between different
metallicity-sensitive indices. This result clearly indicates the importance
of combining several indices in order to distinguish between occasional
abundance effects and contribution of superyoung stars.

To illustrate our findings, we analyse here the behaviour of the galaxy 
with MaNGA identification number 1-211098 (yellow data point in Fig.\,\ref{fig:standard}
and \ref{fig:red}). This object has
a mean luminosity-weighted age $\sim$3\,Gyr (or slightly older), $[$M/H$]\sim$0.0,
$[$Z$_{\rm Mg\emph{b}}$/Z$_{\rm Fe4383}]\sim$0.4, and $[$Z$_{\rm Ca4227}$/Z$_{\rm Fe4383}]$$>$0.4
(as estimated through
the proxy presented in \citealt{Vazdekisetal2001, Vazdekisetal2010}. 
Note only estimates like these ones are enough for the purposes of this work).
In the TiO$_1$ and TiO$_2$ panels,
the corresponding data point is shifted horizontally (towards the right),
consistent with a contribution from a $>$8\,Myr-old and $\leq$30\,Myr-old population.
The panels corresponding to Fe6189 and Ca2 show
that the data point is also shifted towards the right. The Ca2 measurement for this galaxy 
has to be 
taken with caution though, since its extremely large value is due to residual sky contamination 
(as seen and discussed in Fig.\,\ref{fig:spectra} and Sec\,\ref{sec:quantitatively}). This issue
has been considered when analysing this galaxy, as well as the other galaxy with a high value of Ca2.

Continuing with the analysos of  MaNGA 1-211098 (yellow data point), it
is located towards the left in the Ca6162 model 
grid with respect to the standard expectation
(we remark Calcium is slightly enhanced in this galaxy).
All the previously described 
movements agree with this galaxy hosting a significant contribution of 10\,Myr-old stars.\\

We remark that an analogous behaviour is obtained with other red indices
from the same species,
such as the remaining lines from the Calcium Triplet. We have selected here
a subset of red indices, carefully chosen to allow distinction between 
abundance effects and true contribution from recent star bursts as well as to
avoid telluric lines at the redshift of the objects.

\subsection{Constraining the contribution of the very young component}\label{sec:quantitatively}
\begin{figure}
 \vspace{2pt}
 \includegraphics[bb=30 40 320 600, angle=0., width=.45\textwidth]{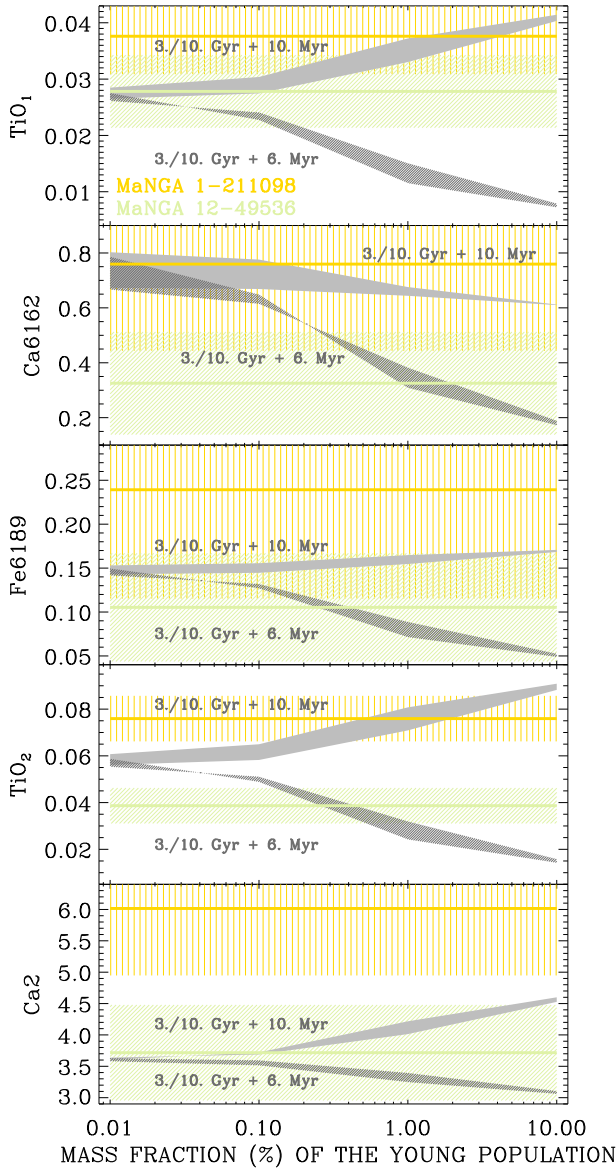}
 \caption{Comparison between the measurements of the red line-strength indices for the galaxies MaNGA
1-211098 (yellow lines) and MaNGA 12-49536 (green lines), and
for combinations of solar-metallicity SSPs including a young burst
(grey regions). 
Indices are organised from the bluest to the reddest
central wavelength (from top to bottom):  TiO$_1$, Ca6162, Fe6189, TiO$_2$, and Ca2.
The grey regions delimit the predictions for an old population (between 3 and 10\,Gyr)
with a certain fraction (as indicated in the horizontal axis) of a 6 or 10\,Myr-old contribution.
The colour vertical and inclined lines account for errors in the index measurements.}
 \label{fig:indicesbursts}
\end{figure}

\begin{figure}
 \vspace{2pt}
 \includegraphics[bb= 35 38 325 765, angle=0., width=.4\textwidth]{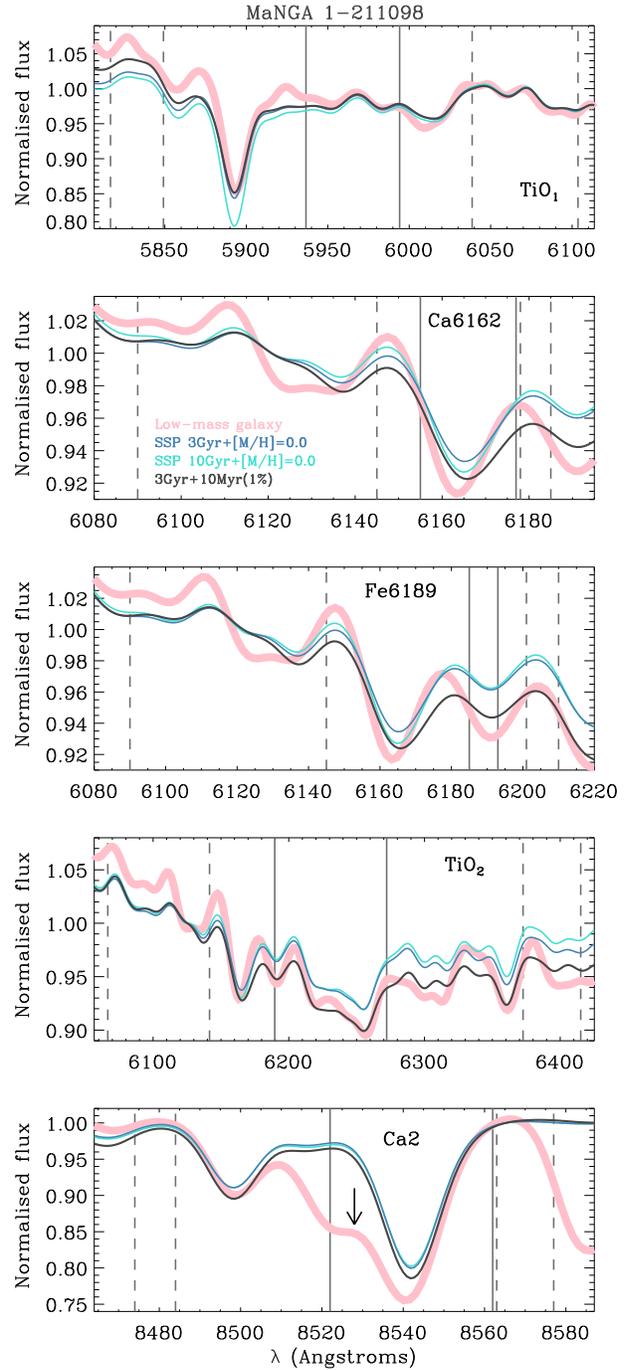}
 \caption{Sections of the stacked spectrum (pink) for the low-mass galaxy MaNGA 1-211098 around
the red absorption lines studied in this paper, as indicated in the panels. The region of the central feature for
the corresponding line indices is indicated with vertical solid lines, while the dashed lines
delimit the blue and red pseudocontinua. The overlapped spectra
show a single stellar population of 3\,Gyr and $[$M/H$]$=0.0 (blue), 
a single stellar population of 10\,Gyr and $[$M/H$]$=0.0 (cyan), and a combination of two
SSPs with solar metallicity: a 3\,Gyr population with a 1\% contribution of 
10\,Myr (dark grey). All spectra have been normalised in the spectral
range covered by both pseudocontinua of each index definition. 
The observed features are better reproduced when a contribution of 10\,Myr-old stars is taken into account.
Note the sky contamination in the region of the Ca2 line affecting this galaxy (highlighted with an arrow,
see text for details).}
 \label{fig:spectra}
\end{figure}

An accurate measurement of the fraction and age of the superyoung population contributing to the eight galaxies
under study is a highly complicated task affected by modelling problems
(inverted behaviour of the metallicity in the superyoung regime, see Sec.\,\ref{sec:discussion}) 
and degenerations 
(e.g., a proper modelling of the star formation history
of the galaxy would be needed). 
Such a detailed analysis is beyond the scope of this
paper, but we aim at showing how some constraints can be provided even 
with the currently available predictions and data.

Figure\,\ref{fig:indicesbursts} shows the ranges of the red metallicity-sensitive indices that are covered by 
an old population ranging from 3 to 10\,Gyr and a young burst that happened either 10 or 6\,Myr ago.
All stars are considered to have solar metallicity. The mass fraction of the superyoung population
varies from 0.01\% (when the effect is almost negligible, as already shown by the 
tracks plotted on Fig.\,\ref{fig:redSSP}) up to 10\%. The fact that the behaviour of the index differences
diverges between young populations older and younger than 8\,Myr illustrates how difficult it would be to 
determine the exact age of the superyoung contribution if a composite history instead of a 
single burst were considered. The detection of such superyoung population is the main focus of this paper.

Figure\,\ref{fig:indicesbursts} compares the old SSP + young burst predictions with the actual measurements
for two galaxies with small contributions from young stellar populations of different ages. The two galaxies
are MaNGA 1-211098 (the yellow data point in Fig.\,\ref{fig:standard} and \ref{fig:red}, used as reference in Sec.\,\ref{sec:red})
and MaNGA 12-49536 (identified with the lightest green colour in Fig.\,\ref{fig:standard} and \ref{fig:red}). 
They both have a mean luminosity-weighted age around 3\,Gyr and solar metallicity, as indicated by the standard indices.
Despite the large error bars affecting the red indices, it is favoured
that MaNGA 1-211098 contains a $>$1\%  mass fraction of a 10\,Myr-old population, whereas
MaNGA 12-49536 contains a $\sim$0.2\% mass fraction of a 6\,Myr-old population.
Once again, the necessity of combining
several features for detecting the superyoung contributions becomes clear with the 
results shown in Fig.\,\ref{fig:indicesbursts}.\\

Figure\,\ref{fig:spectra} shows the spectrum of MaNGA 1-211098, zooming in the spectral
ranges covering the red features of interest. Not only the central feature but also
both red and blue pseudocontinua intervene in the line-strength index computation. 
A comparison with a pure SSP of 3\,Gyr and solar metallicity (close to the standard measurements)
highlights differences
between the galaxy spectrum and the model predictions in those spectral ranges, either in the
features themselves or in the pseudocontinua (all spectra have been normalised in the ranges
covered by both pseudocontinua for each index definition).
Such differences smooth out when a 1\% contribution of a 10\,Myr-old population is added up, as 
previously discussed for this galaxy.
The differences between all models and data in the spectral range of the Calcium Triplet are partly due
to sky lines that were not successfully removed during the MaNGA reduction process, as explained in 
Sec.\,\ref{sec:red}. Such effect has been highlighted in Fig.\,\ref{fig:spectra} and taken into account when 
analysing the signatures of superyoung populations in the sample galaxies.
Once again we note that providing an exact estimate of the amount, metallicity, and age
of the superyoung contribution is not possible due to the several uncertainties affecting this
problem, but this analysis rather provides robust evidence of the presence of such superyoung populations
in the sample low-mass galaxies previously thought to be quenched.

\section{Discussion and Summary}\label{sec:discussion}

We report on the detection of small fractions of superyoung stellar populations in the stacked spectra of 
low-mass galaxies (1$\times$10$^9$$<$\,M/M$_\odot$$<$5$\times$10$^9$) 
thought to be quenched systems. Among the sample of 
28 galaxies under study, eight out of them have $\leq$1\% of their mass in 
$\leq$20\,Myr-old stars. Such contribution has been revealed by the analysis of line-strength indices
in the red regime: from TiO$_1$ to the Calcium Triplet or, equivalently, from 5700\,\AA\ to 
8800\,\AA. This represents a shift toward higher wavelenghts than usually studied in stellar population
analyses.

\subsection{Sensitivity of red indices to very young stellar populations and caveats}

Our results demonstrate the capability 
of red absorption line-strength indices in the range $[$5700,8800$]$\,\AA\
to unveil the presence of superyoung stars in the spectra of external galaxies. 
This procedure is based on the fact that the flux contribution from RSGs
increases in that red regime.
The red supergiant phase lasts short times ($\leq$10\% of the lifetime of a star with mass M$\leq$30\,M$_\odot$),
and is dominant between 8 and 25\,Myr after a stellar burst \citep{Conti1975,Ekstrometal2013}. 
As a result, RSGs
are responsible for the non linearity of the absorption line-strength indices at young ages,
such effect being more pronounced in the red wavelength range than in the standard range.
In Fig,\,\ref{fig:spectra}, a comparison between a solar metallicity, 3\,Gyr-old stellar population with and without a
1\% mass fraction of a superyoung (10\,Myr-old) population is shown. Even within the red $[$5700,8800$]$\AA\ regime,
differences between both spectra in the TiO$_2$ band are more 
significant than in the bluer TiO$_1$ index. Furthermore, those two spectra are more different than the 3\,Gyr-old SSP 
with respect to the 10\,Gyr-old SSP.

Our diagnostics is purely based on the stellar component of the galaxy spectra and therefore
differs from the more common practice of studying the ionised emission lines to detect recent
or ongoing star formation. 
It has the advantage of being sensitive up to 20\,Myr-old stars, whereas
emission in H$\alpha$ is observed only if the bursts are as recent as 10\,Myr or less.
A different stellar approach for detecting recent star formation bursts has been used in 
e.g. \citet[][]{Sanchezetal2019}, \citet{MendezAbreuetal2019b} and \citet{MendezAbreuetal2019c}. Using
full-spectrum-fitting techniques instead of line-strength indices, they measure the amount of stellar 
populations younger than 32\,Myr present in a sample of massive early-type galaxies. These works are however
not sensitive to such tiny fractions of young populations as the ones detected here 
\citep{Bitsakisetal2019}.\\

Some difficulties arise when using the stellar spectra for studying
such little fractions of stars. Deviations of the index measurements from their
standard behaviour in the optical range can be due to 
a number of reasons, such as variations in the abundances of the chemical species and, in addition, 
contamination of telluric and sky emission lines. While this latest effect can be
assessed and even prevented with a careful sample selection, 
our knowledge of chemical abundances in galaxies is still limited but it
will certainly improve in the following years thanks to the effort of several groups.
We note however that this difficulty can be partially overcome by combining
several indices from different species and taking into account the abundant
behaviour observed in the standard range, as done in this work.

We remark again that low-mass galaxies with a relatively old mean luminosity-weighted stellar population
are best suited to perform the kind of study presented in this paper. This is because a small number
of superyoung stars corresponds to a more significant fraction of the low mass of a dwarf than of the mass of 
a giant galaxy. Moreover, the effect of a superyoung population will contrast the most against an old underlying 
population placed at the bottom regions of the index-index diagrams. However, this very same reason
may be behind some of the dispersion observed in the index-index measurements.
After the accumulation of gas that causes the star forming bursts detected in this paper, with a high relative
impact in such low-mass systems, the gas gets heated
and further star formation is therefore prevented. The recent star formation process in these systems
is thus expected to be bursty. On the contrary, the same effect in a massive galaxy would be mild and more generations of 
stellar populations would be born, thus providing a smoother star formation history \citep{diCintioetal2014}.\\

While the detection of the recent starbursts has been unequivocally assessed in the present study, 
a quantitative analysis providing the exact mass fraction, age, and metallicity of the superyoung component
is a very complicated goal, hampered by: i) old-known modelling problems in the superyoung regime that 
prevent the use of specific parameters from the SSPs \citep{LangerandMaeder1995}. 
We particularly refer to \citet{Asadetal2017} for more information on how the inverted trend in the modelling of blue/red
supergiants with respect to observations does not affect our conclusions on the existence of a superyoung
population, but only prevents an accurate determination of its metallicity;
ii) large error bars associated to the
measurement of line-strength indices in the red regime; and iii) the usual uncertainties affecting 
the fitting of multiple stellar populations, since not only the young component but also the 
underlying old population needs to be accurately modelled. We have shown that the
best possible results are reachable when, again, a combination of different species is considered. In particular,
we have made used of TiO$_1$, Ca6162, Fe6189, TiO$_2$, and Ca2.\\

As a main conclusion, we remark the potential of the scarcely-studied red regime for stellar population
studies. More work is needed to better understand the effect of stellar evolution in such wavelength
range and to improve the analysis presented here in this young age range, so it can be extended to more massive galaxies
providing more quantitative results. Having observational data with higher SNR, for which the red indices can be measured with a higher
accuracy and the underlying old population can be better assessed, would certainly benefit this kind of analysis and would
allow a more precise determination of the age and mass fraction of the superyoung population.

\subsection{Implications for the evolution of low-mass galaxies}
During the last 20 years, the research about dwarf galaxies 
in the local Universe has revolved around the idea that they are mostly quenched systems
where the environment might have played an important role at halting their star formation,
many dwarfs belonging to high-density regions such as groups or clusters
\citep[see][and the remaining references in Sec.\,\ref{sec:intro}]{HeldandMould1994,Michielsenetal2008,Sybilskaetal2018}.
Such quenched nature is not opposed to the variety of mean luminosity-weighted age measurements
obtained for these systems, ranging from 1 to 10\,Gyr \citep[e.g.][]{Senetal2018,Sybilskaetal2017}. 
Some dwarf ellipticals have indeed been found to host a relatively young population ($\leq$5\,Gyr-old)
over an underlying old system, thus resulting in a not so old mean age. 
However, and as expected for quenched objects, they do not show signs of very recent star formation.

Our sample galaxies had been purposely selected to include only quenched systems \citep{Pennyetal2016}. 
Moreover, they all lie close to a bright, more massive neighbour with M$>$5$\times$10$^{10}$M$_\odot$ and thus
\citet{Pennyetal2016} pointed out interactions are most likely the cause of their quenching. 
The fact that we detect little percentages of stars formed less than
20\,Myr ago in eight of our galaxies indicates notwithstanding that there is some gas present in them.

\citet{Rysetal2015} studied the star formation histories of a sample of 12 dwarf ellipticals in the 
Virgo cluster and in the field. They concluded that some dwarf ellipticals have gone through
an environmental harassment process that is able to quench star formation but does not fully remove the gas.
The gas reservoir is kept in the galaxies until more recent tidal interactions 
drive it to the central galaxy regions, where it accumulates and may form stars again.
Interestingly, \citet{SalvadorRusinoletal2020}, who also detect $\leq$1\% mass fractions of young populations
but in massive galaxies instead of in dwarfs, suggested that the gas responsible for recent starbursts
may come from late stages of stellar evolution happening inside the galaxies.
Such scenarios, that picture an internal origin for the gas, are 
in agreement with our findings for the eight dwarf galaxies with superyoung stellar populations. 
Note however that a late accretion of external gas
is also a plausible scenario. This possibility would allow for a very efficient gas stripping during the 
quenching phase of the galaxies \citep{LinandFaber1983}.
In an attempt to shed some light over this circumstance, we have 
looked for trends between our sample galaxies and the properties of their massive neighbours 
(mass, brightness, and distance), finding no particular relations with the individuals showing
a superyoung starburst.\\

Given the fact that we have stacked the integral-field spectra of each galaxy in order to increase
the signal-to-noise ratio, we cannot constrain where in the galaxy the recent star formation is taking
place. Attending to the aforementioned scenario proposed by \citet{Rysetal2015}, star formation would be triggered
again in the galaxy centre. This is in agreement with the results from \citet{Kolevaetal2009b}, 
who found that the youngest stellar populations in dwarfs lay more concentrated than the oldest stars.
More recently, \citet{Zhouetal2020} found the opposite result: stars in the outer regions of dwarfs are younger
than those in the centre.
Note that, although \citet{Zhouetal2020} also withdrew their sample from the MaNGA survey as we do,
they work with lower mass individuals with M$<$10$^9$M$_\odot$. It is also plausible that more than one mechanism
are triggering new star formation in this kind of galaxies and that the main driver depends on properties such as
mass, thus explaining the differences found between \citet{Kolevaetal2009b} and \cite{Zhouetal2020}.

We have explored the location of the stacked spaxels in our low-mass targets (driven by an accurate data reduction
and stellar kinematics analysis, as explained in Sec.\,\ref{sec:sample}), finding that they 
cover the galaxy homogeneously for four individuals whereas they are mostly placed in the outer regions
of the remaining four. We remind that we normalised the spectra before stacking and therefore 
the external parts contribute more than the 
central galaxy regions in each final single spectrum. 
This could support the result from \citet{Zhouetal2020} and the fact that the gas has an external
origin, although we remark that we cannot fully discard the reservoir scenario from \citet{Rysetal2015} since star formation
may as well being happening in the centre of our galaxies.

\citet{Pennyetal2018} discovered hints of ionised gas in 14 out of 69 quenched low-mass galaxies from 
the MaNGA survey, none of them in common with our set of galaxies with a superyoung burst. 
For five of their galaxies, there is a kinematical misalignment between the gaseous and stellar components, 
suggestive of a recent accretion of the gas that it is notwithstanding homogeneously distributed across the galaxies. 
We have furthermore explored the flux of H$\alpha$ in emission as provided by the MaNGA data analysis pipeline
\citep{Westfalletal2019} for the same spaxels used in our stacking, looking for even very slight H$\alpha$ content
in our galaxies with superyoung bursts in comparison with the rest of the sample. We find no trend at all (as a reference,
Table\,\ref{tab:sample} shows the median and standard deviation of H$\alpha$ emission flux in the 
spaxels used in the analysis for the whole sample).
The picture is therefore not clear yet. A proper analysis of the 
star formation histories and environment of our low-mass galaxies is required to provide a robust answer
in this regard. In any case, the absence of H$\alpha$ emission in our superyoung low-mass galaxies 
supports the need for an alternative analysis like the one presented in this paper in order to detect 
such low level of star formation.\\

Finally, we would like to emphasise that our approach of considering two SSP for constraining the youngest stellar
contribution is based on previous works \citep[see e.g.][]{Kolevaetal2009b, Rysetal2015}.
The bulk (more than 99\% in mass) of the stellar populations in our low-mass galaxies is very old,
with mean luminosity-weighted ages $\geq$2\,Gyr. While the detailed star formation histories of the
galaxies more than 2\,Gyr ago are most likely complicated and different among individuals
\citep[extended or composed of several bursts;][]{Rysetal2015}, 
their representation with a single stellar population serves well our purposes
of qualitatively identifying a superyoung starburst.

\section*{Acknowledgments}
We are indebted to Seyda {\c S}en, Reynier Peletier, and Francesco La Barbera 
for original work and discussions that motivated the development
of this project, as well as to the referee for a very thorough revision of our manuscript. 
We acknowledge support from grant AYA2016-77237-C3-1-P 
from the Spanish Ministry of Science and Universities (MCIU). MAB also acknowledges support from
the Severo Ochoa Excellence scheme (SEV-2015-0548).
This work makes use of data from SDSS-IV. Funding for
SDSS-IV has been provided by the Alfred P. Sloan Foundation and Participating Institutions. 
Additional funding towards SDSS-IV has been provided by the US Department of Energy Office of Science. 
SDSS-IV acknowledges support and resources from the Centre for High-Performance Computing at the 
University of Utah.

\section*{Data availability}
This work makes use of data from the MaNGA Public Release 5 (MPL-5) of the SDSS-IV project.
All the datacubes, stellar kinematic and emission lines analyses are available through the 
corresponding webpages. The SDSS web site is www.sdss.org while information about MaNGA can 
be found in https://www.sdss.org/surveys/manga/.

\bibliographystyle{mn2e}
\bibliography{reference}

\bsp

\label{lastpage}

\end{document}